\documentclass[a4paper,11pt]{article}
\usepackage{pos}

\title{The DM approach to semileptonic heavy-to-heavy and heavy-to-light $B$ decays}
\ShortTitle{The DM approach to semileptonic heavy-to-heavy and heavy-to-light $B$ decays}

\author[a]{G. Martinelli}
\author[b]{M. Naviglio}
\author[c]{S. Simula}
\author*[d]{L. Vittorio}

\affiliation[a]{Physics Department and INFN Sezione di Roma La Sapienza,\\
  Piazzale Aldo Moro 5, 00185 Roma, Italy}
  
\affiliation[b]{Dipartimento di Fisica dell'Universit\`a di Pisa and INFN, Sezione di Pisa,\\ Largo Bruno Pontecorvo 3, I-56127 Pisa, Italy}
  
\affiliation[c]{Istituto Nazionale di Fisica Nucleare, Sezione di Roma Tre,\\
Via della Vasca Navale 84, I-00146 Rome, Italy}

\affiliation[d]{LAPTh, Universit\'e Savoie Mont-Blanc and CNRS, Annecy, France}

\emailAdd{guido.martinelli@roma1.infn.it}
\emailAdd{manuel.naviglio@phd.unipi.it}
\emailAdd{silvano.simula@roma3.infn.it}
\emailAdd{ludovico.vittorio@lapth.cnrs.fr}

\abstract{We present the results of the application of the Dispersion Matrix approach to semileptonic heavy-to-heavy and heavy-to-light $B$-meson decays. This method allows to determine the hadronic form factors in a non-perturbative and model-independent way. Starting from the available lattice results at large values of the momentum transfer, we obtain the behaviour of the form factors in their whole kinematical range without introducing any parameterization of their momentum dependence. We will focus on the determination of the Cabibbo-Kobayashi-Maskawa matrix elements $\vert V_{cb} \vert$ and $\vert V_{ub} \vert$ through the analysis of $B \to D^{(*)} \ell \nu$, $B_s \to D_s^{(*)} \ell \nu$, $B \to \pi \ell \nu$ and $B_s \to K \ell \nu$ decays. New theoretical determinations of the Lepton Flavour Universality ratios relevant for these transitions will be also presented, by focusing in particular on the $R(D_{(s)}^{(*)})$ ratios.}

\FullConference{%
The 39th International Symposium on Lattice Field Theory,\\
8th-13th August, 2022,\\
Rheinische Friedrich-Wilhelms-Universität Bonn, Bonn, Germany
}

\begin{document}
\maketitle

\section{Introduction}
\label{intro}

Semileptonic $B$-meson decays are one of the most challenging processes in the phenomenology of flavour physics, as they are affected by two unsolved problems. 

On the one hand, a non-negligible tension exists between the inclusive and the exclusive determinations of the CKM matrix element $\vert V_{cb}\vert$. This discrepancy is also knonw as the $\vert V_{cb} \vert$ \emph{puzzle}. According to the FLAG Review 2021 \cite{FLAG21}, we have a $\sim2.8\sigma$ tension between the exclusive estimate and the inclusive one, namely
\begin{equation}
\label{VcbFLAG21}
\vert V_{cb} \vert_{\rm excl} \times 10^3 = 39.36(68),\,\,\,\,\,\,\,\,\,\,\,\,\vert V_{cb} \vert_{\rm incl} \times 10^3 = 42.00(65).
\end{equation}
Two new estimates of the inclusive value have also recently appeared, $i.e.$ $\vert V_{cb} \vert_{\rm incl} \times 10^3=42.16(50)$ \cite{Bordone:2021oof} and $\vert V_{cb} \vert_{\rm incl} \times 10^3=41.69(63)$ \cite{Bernlochner:2022ucr}, which are compatible with the inclusive FLAG value in Eq.\,(\ref{VcbFLAG21}) and corroborate its robustness. 

On the other hand, we have a strong tension between the theoretical values and the measurements of the $R(D^{(*)})$ ratios, which are defined as 
\begin{equation}
\label{RDdef}
R(D^{(*)}) \equiv \frac{\Gamma(B \to D^{(*)} \tau \nu_{\tau})}{\Gamma(B \to D^{(*)} \ell \nu_{\ell})},
\end{equation}
where $\ell$ denotes a light lepton. The HFLAV Collaboration \cite{HFLAV} has recently computed the world averages of the available measurements of $R(D^{(*)})$ and of their SM theoretical predictions. To be more specific, we have 
\begin{equation}
\label{RDHFLAV}
R(D)_{\rm SM} = 0.298 \pm 0.004,\,\,\,R(D)_{\rm exp} = 0.339 \pm 0.026 \pm 0.014
\end{equation}
for the $B \to D$ case and 
\begin{equation}
\label{RDstHFLAV}
R(D^*)_{\rm SM} = 0.254 \pm 0.005,\,\,\,R(D^*)_{\rm exp} = 0.295 \pm 0.010 \pm 0.010
\end{equation}
for the $B \to D^*$ one. As clearly stated by HFLAV Collaboration, the world averages of the measurements of $R(D)$ and $R(D^*)$ exceed the corresponding SM expectations by 1.4$\sigma$ and 2.8$\sigma$, respectively. If the experimental correlation between these two quantities, namely $\rho=-0.38$, is also taken into account, then the resulting difference with the SM predictions is increased at the 3.3$\sigma$ level.

\section{The Dispersion Matrix (DM) approach}
\label{Section2}
Let us focus on $B \to D^{(*)} \ell \nu$ decays for massless leptons (namely $\ell = e, \mu$). In case of production of a \emph{pseudoscalar} meson, $i.e.$ the $B \to D \ell \nu$ case, the differential decay width reads
\begin{equation}
\label{finaldiff333}
\frac{d\Gamma}{dq^2}=\frac{G_F^2}{24\pi^3} \eta_{EW}^2 \vert V_{cb} \vert^2 \vert \vec{p}_{D}\vert^3  \vert f^+(q^2) \vert^2,
\end{equation}
where $\vec{p}_{D}$ represents the 3-momentum of the produced $D$ meson. In case of production of a \emph{vector} meson, $i.e.$ the $B \to D^* \ell \nu$ case, the expression of the differential decay width is much more complicated, $i.e.$
\begin{equation}
\begin{aligned}
\label{finaldiff333BDst}
\frac{d\Gamma(B \rightarrow D^{*}(\rightarrow D\pi) \ell \nu)}{dw d\cos \theta_{\ell} d\cos \theta_v d\chi}& =\frac{3}{4}\frac{G_F^2 }{(4\pi)^4} \eta_{EW}^2 \vert V_{cb} \vert^ 2 m_B^3 r^2 \sqrt{w^2-1}  \\
& \times (1+r^2-2rw) \{ (1-\cos \theta_{\ell} )^2 \sin^2 \theta_v \vert H_{+} \vert^2 \\
& + (1+\cos \theta_{\ell} )^2\sin^2 \theta_v \vert H_{-} \vert^2+ 4 \sin^2 \theta_{\ell}\cos^2 \theta_v\vert H_{0} \vert^2 \\
& - 2 \sin^2 \theta_{\ell}\sin^2 \theta_v \cos 2\chi  H_{+}  H_{-} \\
& - 4 \sin \theta_{\ell} (1-\cos \theta_{\ell} ) \sin\theta_v\cos\theta_v\cos\chi H_{+}  H_{0}\\
& + 4 \sin \theta_{\ell} (1+\cos \theta_{\ell} ) \sin\theta_v\cos\theta_v\cos\chi H_{-}  H_{0}\},
\end{aligned}
\end{equation}
where we have defined $r \equiv m_{D^*}/m_B$ and we have introduced the helicity amplitudes 
\begin{equation}
\label{helampl}
H_0(w) = \frac{\mathcal{F}_1(w)}{\sqrt{m_B^2+m_D^2-2m_Bm_Dw}},\,\,\,\,\,H_{\pm}(w) = f(w) \mp m_B m_{D^*} \sqrt{w^2-1}\,g(w).
\end{equation}
Since the $D^*$ meson strongly decays into a $D \pi$ pair, we have to define the so-called helicity angles, which are called $\theta_l, \theta_v, \chi$ in Eq.(\ref{finaldiff333BDst}).

The hadronic FFs are the quantities $f^+(q^2)$ in Eq.\,(\ref{finaldiff333}) and $f(w),\,g(w),\,\mathcal{F}_1(w)$ in Eq.\,(\ref{finaldiff333BDst}). In what follows, we will refer equivalently to the momentum transfer $q^2$ or to the recoil $w$, since they are related by the following 1-to-1 correspondence
\begin{equation}
q^2(w)= m_B^2+m_{D^{(*)}}^2 - 2 m_B m_{D^{(*)}} w.
\end{equation}
Let us finally stress again that this is a simplified picture, since we are assuming a massless produced lepton. In fact, for massive leptons ($\ell = \tau$) one should consider also the FFs $f_0(q^2)$ for semileptonic $B \to D$ decays and $P_1(w)$ for semileptonic $B \to D^*$ ones.

Now, the goal of this proceedings is to describe the FFs entering in semileptonic heavy-to-heavy and heavy-to-light $B$-meson decays by using the novel Dispersion Matrix (DM) method \cite{DiCarlo:2021dzg}. The DM method, in fact, allows us to study the FFs in a non-perturbative and model-independent way. To be more specific, starting from the available LQCD computations of the FFs at high momentum transfer (or, equivalently, at low recoil), we can extrapolate their behaviour in the opposite kinematical region without assuming any functional dependence of the FFs on $q^2$ (or, equivalently, on $w$) and using only non-perturbative inputs. From the mathematical point of view, the starting point is to focus on one FF, let us call it generically $F$, and then build up the matrix
\begin{equation}
\label{eq:Delta2}
\mathbf{M} = \left( 
\begin{tabular}{ccccc}
   $\chi$ & $\phi F$ & $\phi_1 F_1$ & $...$ & $\phi_N F_N$ \\[2mm] 
   $\phi F$ & $\frac{1}{1 - z^2}$ & $\frac{1}{1 - z z_1}$ & $...$ & $\frac{1}{1 - z z_N}$ \\[2mm]
   $\phi_1 F_1$ & $\frac{1}{1 - z_1 z}$  & $\frac{1}{1 - z_1^2}$ & $...$ & $\frac{1}{1 - z_1 z_N}$ \\[2mm]
   $... $  & $...$ & $...$ & $...$ & $...$ \\[2mm]
   $\phi_N F_N$ & $\frac{1}{1 - z_N z}$ & $\frac{1}{1 - z_N z_1}$ & $...$ & $\frac{1}{1 - z_N^2}$
\end{tabular}
\right),
\end{equation}
where we have introduced the conformal variable $z$ defined as 
\begin{equation}
\label{conf}
z(t) = \frac{\sqrt{t_+ -t} - \sqrt{t_+ - t_-}}{\sqrt{t_+ -t} + \sqrt{t_+ - t_-}},\,\,\,\,\,\,\,\,\,\,\,\,\,\,\,\,\,\,\,\,\,t_{\pm}=(m_B \pm m_{D^{(*)}})^2
\end{equation}
or, equivalently, as
\begin{equation}
z(w)=\frac{\sqrt{w+1}-\sqrt{2}}{\sqrt{w+1}+\sqrt{2}}.
\end{equation}
In Eq.\,(\ref{eq:Delta2}) $\phi_i F_i \equiv \phi(z_i) F(z_i)$ (with $i = 1, 2, ... N$) are the known values of the quantity $\phi(z) F(z)$ corresponding to the values $z_i$ at which the FFs have been computed on the lattice. The kinematical functions $\phi(z)$ have an expression which is specific to each of the hadronic FFs \cite{paperoIII}. Finally, the susceptibility $\chi(q_0^2)$ is related to the derivative with respect to $q_0^2$ of  the Fourier transform of suitable Green functions of bilinear quark operators and follows from the dispersion relation associated to a particular spin-parity quantum channel. They have been computed for the first time on the lattice in \cite{paperoII, paperoV} for $b \to c$ and $b \to u$ quark transitions, respectively, by setting $q_0^2=0$. At this point, one can demonstrate from first principles that $\det \mathbf{M} \geq 0$. The positivity of the determinant, which we will refer to as \emph{unitarity filter} hereafter, allows thus to compute the lower and the upper bounds of the generic FF $F$ for each generic value of $z$, $i.e.$
\begin{equation}
F_{\rm lo}(z) \leq F(z) \leq F_{\rm up}(z).
\end{equation}
The explicit expressions of $F_{\rm lo,up}(z)$ can be found in \cite{DiCarlo:2021dzg}, so that we can rephrase the above condition as
\begin{equation}
   \beta - \sqrt{\gamma} \leq  F(z) \leq \beta + \sqrt{\gamma} ~ , ~
    \label{eq:bounds}
\end{equation} 
where (after some algebraic manipulations)
\begin{equation*}
      \beta = \frac{1}{\phi(z)d(z)} \sum_{j = 1}^N f_j \phi_j d_j \frac{1 - z_j^2}{z_0 - z_j},\,\,\,\,\,\,\,\,\,\, \gamma  =   \frac{1}{1 - z_0^2} \frac{1}{\phi(z)^2 d(z)^2} \left( \chi - \overline{\chi} \right),   
\end{equation*} 
\begin{equation*}
       \overline{\chi}  =  \sum_{i, j = 1}^N f_i f_j \phi_i d_i \phi_j d_j \frac{(1 - z_i^2) (1 - z_j^2)}{1 - z_i z_j}
\end{equation*}
where $d(z),\,d_i$ are kinematical functions. Unitarity is satisfied only when $\gamma \geq 0$, which implies $\chi \geq \overline{\chi}$. One can show that the values of $\beta$ and $\gamma$ depend on $z$, while this is not the case for $\overline{\chi}$. In fact, it depends only on the set of input data. Consequently, the unitarity condition $\chi \geq \overline{\chi}$ does not depend on $z$.

In what follows, we will study in detail an explicit example, namely the application of the DM method to $B \to \pi \ell \nu$ transitions. Then, we will give an overview of all the results relevant for phenomenology obtained so far, enlarging the discussion to both $b \to u$ and $b \to c$ quark transitions.

\section{The DM application to semileptonic $B \to \pi \ell \nu$ decays}

Let us discuss, as an instructive example, the application of the Dispersive Matrix method to the $B \to \pi \ell \nu$ decays \cite{paperoV}. Since the $\pi$ meson is a pseudoscalar one, the formalism is completely analogous to the one characterizing $B \to D \ell \nu$ decays, which has been described in Section \ref{Section2}. Thus, $B \to \pi \ell \nu$ transitions are characterized by two FFs, which we will call $f_+^{\pi}(q^2),\,f_0^{\pi}(q^2)$. These FFs have been studied by the RBC/UKQCD\,\cite{Flynn:2015mha} and the FNAL/MILC\,\cite{Lattice:2015tia} Collaborations. Note that, very recently, the results of a new computation of the hadronic FFs have been also published by the JLQCD Collaboration \cite{Colquhoun:2022atw}. We have planned to include these data in a future analysis, together with the new results that are going to be published by the RBC/UKQCD Collaboration \cite{Flynn:2021ttz}.

Now, the lattice computations of the FFs $f_+^{\pi}(q^2),\,f_0^{\pi}(q^2)$ are available in the large-$q^2$ region. To be more specific, the authors of Ref.\,\cite{Flynn:2015mha} provide synthetic LQCD values of the FFs (together with their statistical and systematic correlations) at $q^2 = \{ 19.0, 22.6, 25.1 \}$ GeV$^2$. In \cite{Lattice:2015tia}, instead, only the results of BCL fits \cite{BCL} of the FFs extrapolated to the continuum limit and to the physical pion point are available. Thus, from the marginalized BCL coefficients we evaluate the mean values, uncertainties and correlations of the FFs at the same three values of $q^2$ given in Ref.\,\cite{Flynn:2015mha}. 

\begin{figure}
\centering
\includegraphics[width=.9\textwidth]{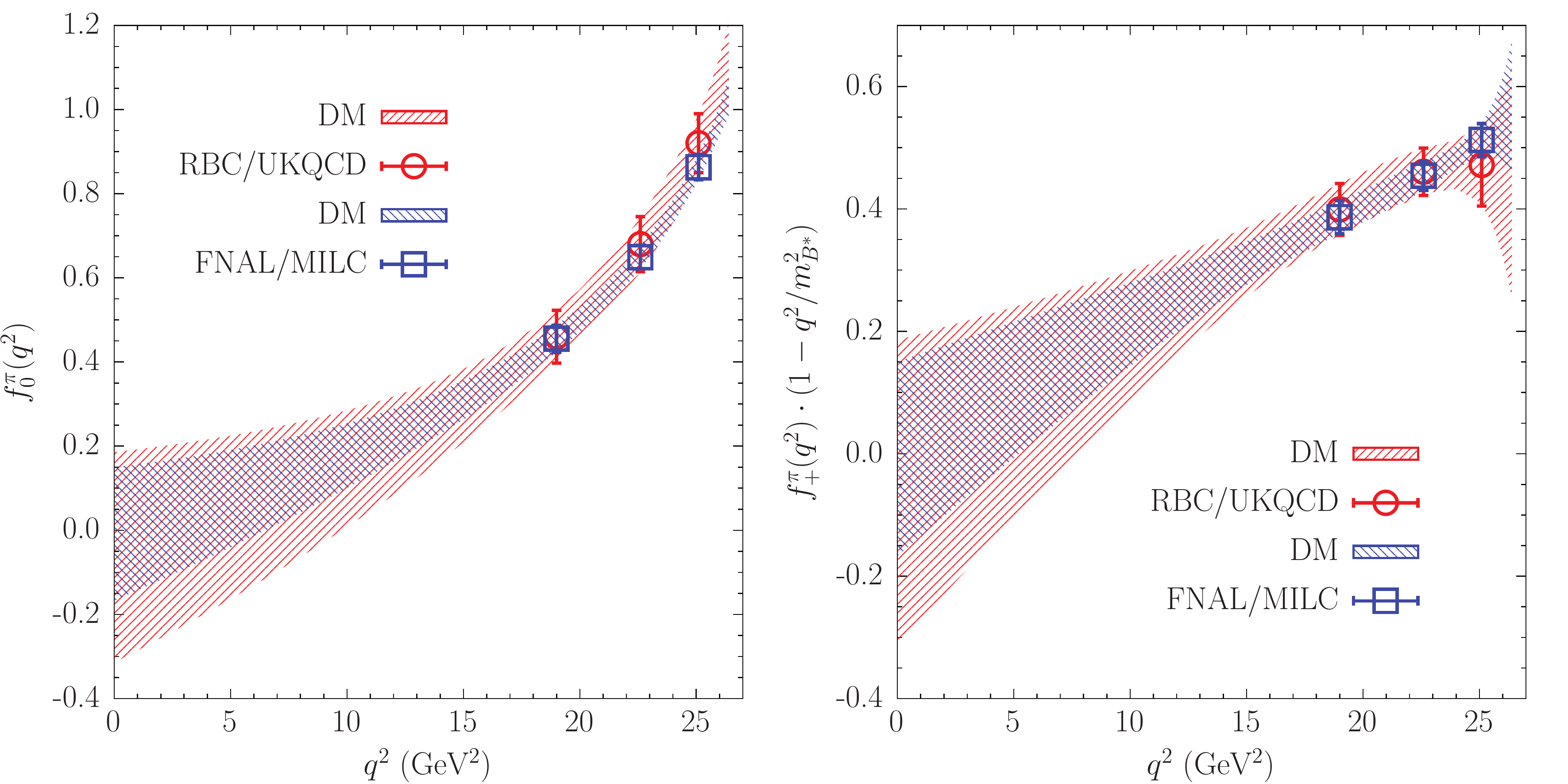}
\caption{\small The scalar $f_0^\pi(q^2)$ (left panel) and vector $f_+^\pi(q^2)$ (right panel) FFs entering the semileptonic $B \to \pi \ell \nu_\ell$ decays computed by the DM method as a function of the 4-momentum transfer $q^2$ using the LQCD inputs from RBC/UKQCD (red points) and FNAL/MILC (blue squares) Collaborations. In the right panel, the vector FF is multiplied by the factor $(1 - q^2 / m_{B^*}^2)$ with $m_{B^*} = 5.325$ GeV.\hspace*{\fill}}
\label{FFsBpi}
\end{figure}

In Fig.\,\ref{FFsBpi} we show the red (blue) DM bands coming from the RBC/UKQCD (FNAL/MILC) data, respectively. In principle, when one implements the BCL fits the mean value and the uncertainty of the FFs value extrapolated at zero momentum transfer are not stable under variation of the truncation order of the series expansion of the FFs. On the contrary, the DM method is completely independent of this issue, since no approximation due to the truncation of a series expansion is present. In other words, we argue that the DM method is equivalent to the results of all possible (BCL) fits which satisfy unitarity and, at the same time, reproduce exactly the input data. This property is particularly useful in $B \to \pi \ell \nu$ decays, since here we have a long extrapolation in $q^2$.

\begin{figure}
\centering
\includegraphics[width=.7\textwidth]{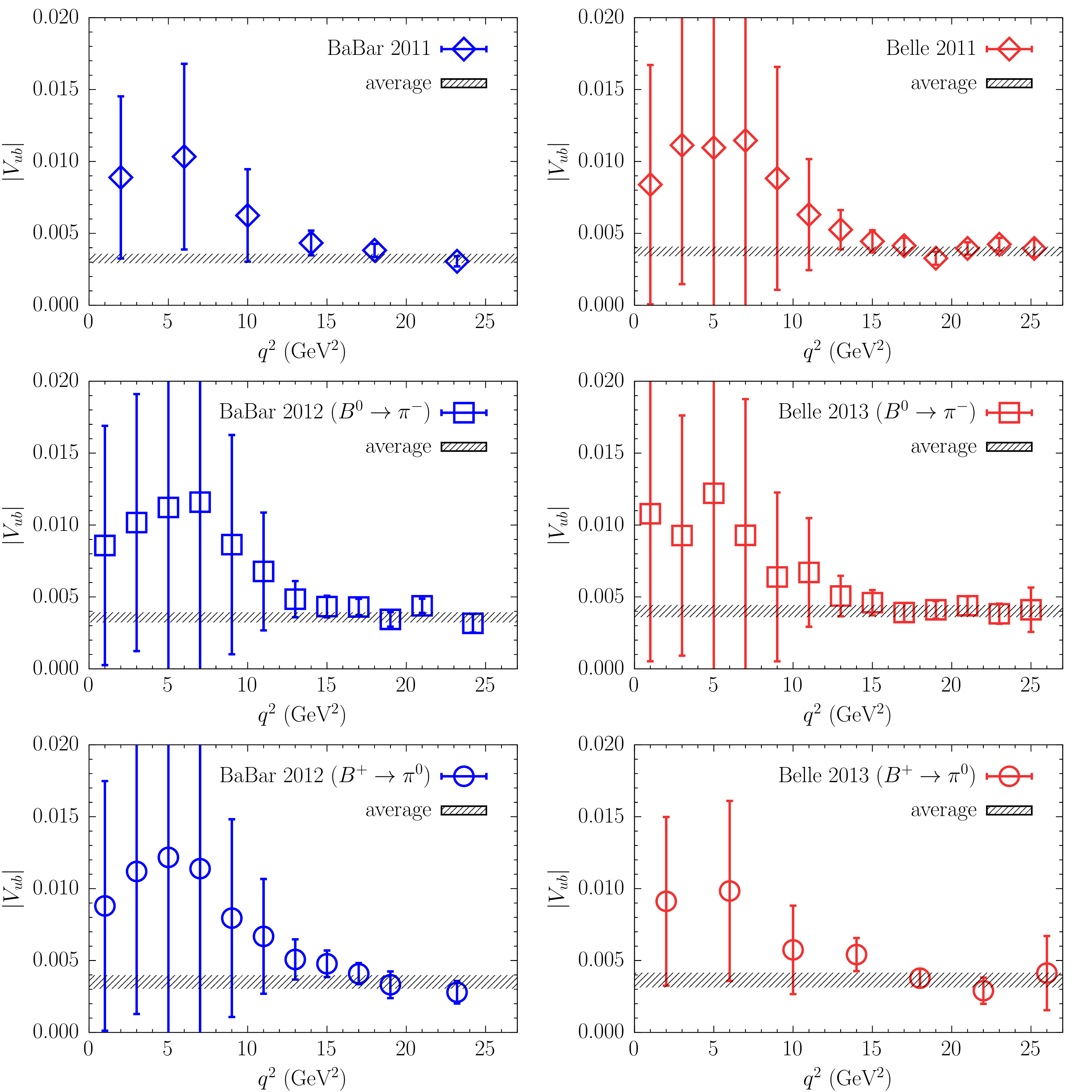}
\caption{\it \small Bin-per-bin estimates of $\vert V_{ub} \vert$ for each of the experiments of Refs.\,\cite{delAmoSanchez:2010af, Ha:2010rf, Lees:2012vv, Sibidanov:2013rkk}, which are specified in the insets of the panels as a function of $q^2$. The theoretical DM bands of the FFs correspond to the use of the combination of the RBC/UKQCD and FNAL/MILC data as inputs. The black dashed bands represent the correlated weighted averages for each experiment.}
\label{VubBpifig}
\end{figure}

For the extraction of the CKM matrix element we compute bin-per-bin values of $\vert V_{ub} \vert$ for each $q^2$-bin of each available experiment. Several experiments \cite{delAmoSanchez:2010af, Ha:2010rf, Lees:2012vv, Sibidanov:2013rkk} have measured the differential branching fractions of the semileptonic $B \to \pi$ transition. To be more specific, we evaluate the CKM matrix element for the $n$-th experiment ($n = 1, \ldots, 6$ for the semileptonic $B \to \pi$ decays) through the expressions 
\begin{equation}
\label{muVcbfinal}
\vert V_{cb} \vert = \frac{\sum_{i,j=1}^{\hat{N}_n} (\mathbf{C}^{-1})_{ij} \vert V_{cb} \vert_j}{\sum_{i,j=1}^{\hat{N}_n} (\mathbf{C}^{-1})_{ij}},\,\,\,\,\,\,\,\,\,\,\,\,\sigma^2_{\vert V_{cb} \vert} = \frac{1}{\sum_{i,j=1}^{\hat{N}_n} (\mathbf{C}^{-1})_{ij}},
\end{equation}
where $\hat{N}_n$ is the number of the bins associated to the $n$-th experiment. Our results are shown in Fig.\,\ref{VubBpifig} when one uses the combination of the RBC/UKQCD and FNAL/MILC data as inputs of the DM method.

Our final result for $\vert V_{ub} \vert$ is evaluated making use of the averaging procedure given by the formul\ae\, ($N=6$ in our case)
\begin{equation}
\label{sigma28}
\mu_{x} = \frac{1}{N} \sum_{k=1}^N x_k,\,\,\,\,\,\,\,\,\,\,\,\,\,\,\,\,\,\,\, \sigma^2_{x} = \frac{1}{N} \sum_{k=1}^N \sigma_k^2 + \frac{1}{N} \sum_{k=1}^N(x_k-\mu_{x})^2
\end{equation}
and reads
\begin{equation}
\label{VubLASTCOMB}
\vert V_{ub} \vert^{B\pi} \cdot 10^{3}  =  3.62 \pm 0.47.
\end{equation}
Let us mention here that we are currently investigating new strategies to improve our precision on the determination of $\vert V_{ub} \vert$ within the DM approach. Some results can be found in \cite{CKM21}, where our improved determination of the CKM matrix element $\vert V_{ub} \vert$ from semileptonic $B \to \pi$ decays reads 
\begin{equation}
\label{Vubimpr}
\vert V_{ub} \vert^{B\pi}_{\rm impr} \cdot 10^{3}  =  3.88 \pm 0.32.
\end{equation}

\begin{table}
\centering
\caption{Numerical values of the CKM matrix elements $\vert V_{cb} \vert$ and $\vert V_{ub} \vert$ plotted in Figure \ref{Summary}.}
\label{tab1}       
\begin{tabular}{c|c|c|c|l|cc}
& Decay channel & DM values & FLAG '21 & Inclusive & UTfit '22\\\hline
$\vert V_{cb} \vert \times 10^3$ & $B_{(s)} \to D_{(s)}^{(*)}$ & 41.2 (8) & 39.48 (68) & 42.16 (50) & 41.27 (89) \\\hline
$\vert V_{ub} \vert \times 10^3$ & $B_{(s)} \to \pi(K)$  & 3.85 (27) & 3.63 (14) & 4.13 (26) & 3.77 (22) \\
\end{tabular}
\end{table}

\begin{table}
\centering
\caption{Numerical values of the LFU observables relevant for $b \to c$ quark transitions, which are plotted in Figure \ref{Summary}.}
\label{tab2}       
\begin{tabular}{c|c|c|c}
& DM values & HFLAV '21 (exp) & HFLAV '21 (SM)\\\hline
$R(D)$   & 0.296 (8) & 0.339 (26) (14) & 0.299 (3) \\\hline
$R(D^*)$ & 0.275(8) & 0.295 (10) (10) & 0.254 (5) \\\hline
$R(D_s)$ & 0.298 (5) & --- & ---\\\hline
$R(D_s^*)$ & 0.250 (6) & --- & ---\\
\end{tabular}
\end{table}

\begin{table}
\centering
\caption{Numerical values of the LFU observables relevant for $b \to u$ quark transitions. There is only one available measurement of $R_{\pi}^{\tau/\mu}$ by the Belle Collaboration  \cite{Hamer:2015jsa}, while $R_{K}^{\tau/\mu}$ has not been measured yet.}
\label{tab3}       
\begin{tabular}{c|c|c}
& DM values & Measurement\\\hline
$R_{\pi}^{\tau/\mu}$   & 0.793 (118) & 1.05 (51) \\\hline
$R_{K}^{\tau/\mu}$   & 0.755 (138) & ---  \\
\end{tabular}
\end{table}

\section{Conclusions}

\begin{figure}
\centering
\includegraphics[width=1.\textwidth]{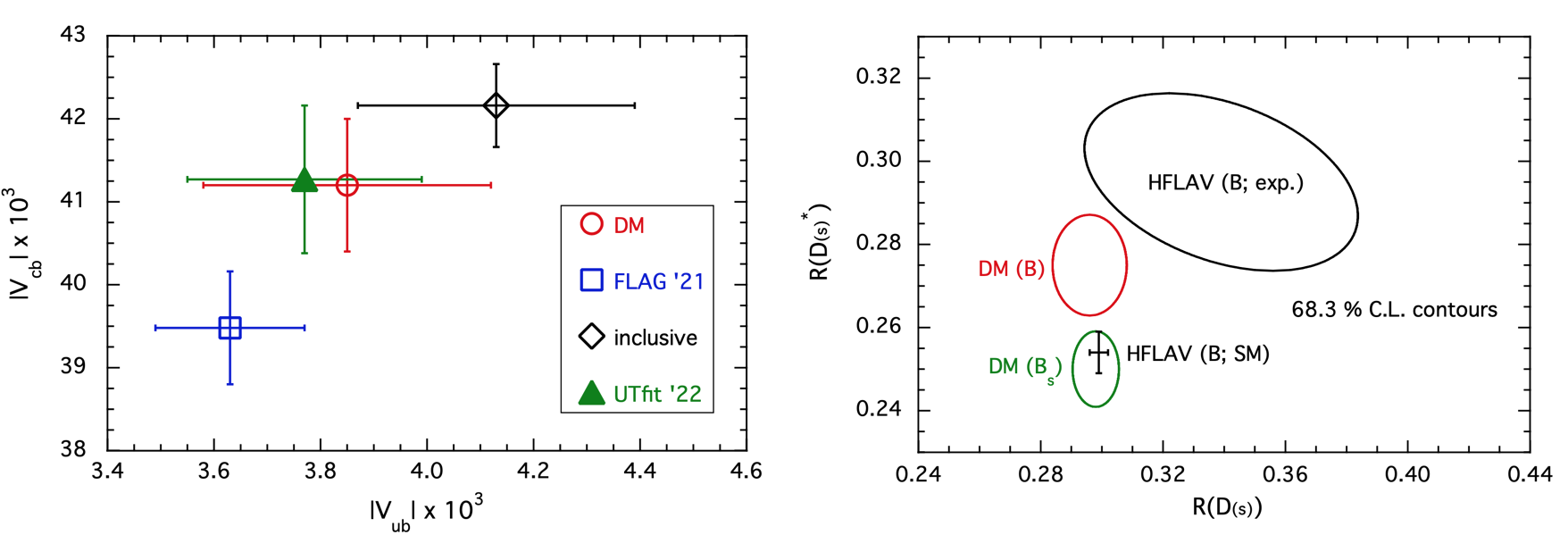}
\caption{\it \small Summary of all the DM results. \textbf{Left panel:} $\vert V_{cb} \vert$ vs $\vert V_{ub} \vert$ correlation plot. The numerical values corresponding to the various entries (Dispersive Matrix estimates, FLAG 2021 values \cite{FLAG21}, inclusive values \cite{Bordone:2021oof, PDG}, UTfit indirect determinations \cite{Bona:2022xnf, Bona:2022zhn}) can be found in Table \ref{tab1}. \textbf{Right panel:} $R(D_{(s)}^*)$ vs $R(D_{(s)})$ correlation plot. The numerical values corresponding to the various entries (Dispersive Matrix estimates, HFLAV experimental average, HFLAV SM average) can be found in Table \ref{tab2}.}
\label{Summary}
\end{figure}

In this contribution we have reviewed the main properties of the Dispersion Matrix approach, which is an interesting tool to implement unitarity and LQCD calculations in the analysis of exclusive charged-current semileptonic decays of mesons and baryons. In Figure \ref{Summary} we have condensed the results obtained so far from the application of the DM method to the semileptonic $B \to D^{(*)}$ \cite{paperoIII, EPJC}, $B_s \to D_s^{(*)}$ \cite{BsDs}, $B \to \pi$ and $B_s \to K$ \cite{paperoV} decays. The DM values of the CKM matrix elements in the left panel represent the averages of all the DM determination of $\vert V_{cb} \vert$ and $\vert V_{ub} \vert$ from the various decay channels, which are also presented in Table \ref{tab1}. For both the CKM matrix elements, the DM determinations are compatible with the corresponding inclusive values within the $1\sigma$ level. Furthermore, the DM values are practically identical to the indirect determinations coming from the latest analysis by the UTfit Collaboration \cite{Bona:2022xnf, Bona:2022zhn}. The values of the LFU observables relevant for semileptonic $B \to D^{(*)}$ and $B_s \to D_s^{(*)}$ decays (that can be found in Table \ref{tab2}) are, instead, shown in the right panel, together with the experimental average and the SM one by HFLAV. By using the FNAL/MILC computations of the FFs for the $B \to D^*$ channel, we can state that the tension between theoretical expectations and measurements of $R(D^{(*)})$ is reduced. Finally, in Table \ref{tab3} the numerical DM estimates of the LFU observables related to the $b \to u$ quark transitions are also presented.

\bibliographystyle{JHEP}
\bibliography{notes_biblio}

\end{document}